\def\rhoo{\rho_{01}}
\def\Rin{\varpi_{0,{\rm in}}}
\def\Rout{\varpi_{0,{\rm out}}}
\def\krho{{k_\rho}}
\def\AU{{\rm AU}}
\def\pc{{\rm pc}}
\def\cm{{\rm cm}}
\def\Msun{M_{\odot}}
\def\nH{{n_{\rm H}}}
\def\pw{p_w}
\def\etain{{\eta_{\rm in}}}

\documentstyle[emulateapj,epsfig,flushrt]{article}
 \def\figcaption[#1]#2{\begin{figure}%
         \begin{center} %
         \epsfig{file=#1}%
         \end{center} %
         \vskip 1in %
         \def\@captype{figure} %
         \caption{#2}%
         \end{figure}}

\begin{document}
%
\title{Bipolar molecular outflows driven by hydromagnetic protostellar winds}
\author{Christopher D. Matzner and Christopher F. McKee}
\affil{Departments of Physics and Astronomy, University of California
at Berkeley}
\authoraddr{601 Campbell Hall, U.C. Berkeley, CA 94720}
\authoremail{matzner@astron.berkeley.edu, cmckee@mckee.berkeley.edu}
\begin{abstract} 
We demonstrate that magnetically-collimated protostellar winds will
sweep ambient material into thin, radiative, momentum-conserving
shells whose features reproduce those commonly observed in
bipolar molecular outflows. We find the typical position-velocity and
mass-velocity relations to occur in outflows in a wide variety 
of ambient density
distributions, regardless of the time histories of their driving winds.
\end{abstract} 

\keywords{hydrodynamics --- ISM: jets and outflows --- stars:
formation --- stars: mass loss}

\section{Introduction} \label{S:intro}

	As they form, young stars emit powerful winds whose mechanical
luminosity amounts to a fair fraction of the stars' binding
energy. These winds are often observed as jets, and it is generally
believed that a toroidal magnetic field is responsible for their
collimation (Benford 1978, Blandford \& Payne 1982).  When a
protostellar wind strikes the ambient medium, a bipolar molecular
outflow is produced. These outflows have been credited with the
support (\cite{NS80}, \cite{M89}, \cite{BM96}) as well as the
disruption (e.g., \cite{BRLB99}) of molecular clouds and the dense
clumps within them that produce star clusters. We seek a model of
protostellar outflows that is sufficiently detailed to permit a
quantitative study of both their supportive and destructive roles in
the lives of their parent clouds.

Bipolar molecular outflows display a number of common features that
must be reproduced in any viable model. As summarized by Lada \& Fich
(1996), these include a nearly linear position-velocity relation (a
``Hubble law'') and a mass-velocity relation $dM/dv\propto v^{\Gamma}$
with $\Gamma \simeq -1.8$. It is debated whether these features result
primarily from turbulent mixing in the region affected by the jet
(e.g., \cite{Ragaea93}, \cite{S94}), or from the dynamics of a shocked
shell bounding the wind cocoon (e.g., \cite{Shuea91}, Masson \&
Chernin 1992, 1993). In this paper we investigate shells of ambient
material set into motion by hydromagnetic protostellar winds, and
demonstrate that these naturally produce the primary characteristics
of observed outflows. (We do not address traces of high-velocity
molecular gas sometimes found within these shells.)

Our analysis of the wind force distribution (\S \ref{S:windforce})
follows Shu et al. (1995) and the suggestion made by Ostriker (1997);
our investigation of shell motion (\S \ref{S:shells}) follows Shu et
al. (1991), Masson \& Chernin (1993), and Li \& Shu (1996). We
generalize these results and show (\S \ref{S:obs}), contrary to the
conclusions of Masson \& Chernin, that a combined model can reproduce
the observed mass-velocity relationship as well as the
position-velocity law.

\section{Force distributions from disk winds and X-winds} \label{S:windforce}

We seek the distribution of wind momentum flux on scales of the dense
clumps in molecular clouds, far larger than any scale associated with
an accretion disk whose wind produces the outflow. The angular
distribution of the wind momentum
injection rate $\dot\pw$ at time $t$ can be written as
\begin{equation}\label{eq:ptildeDef}
\frac{d\dot\pw(t)}{d\Omega} = r^2\rho_w v_w^2\equiv
	\frac{\dot\pw(t)}{4\pi} P(\mu), 
\end{equation}
where $\mu = \cos\theta$ labels directions from the outflow axis.
We wish to determine $P(\mu)$, the normalized force distribution, 
which we assume is constant in time. 

	Shu et al (1995) have shown that $\rho_w\propto
1/(r\sin\theta)^2$ is a good approximation for X-winds.  Since the
wind velocity $v_w$ is approximately the same on different streamlines
in this model, it follows that $\rho_w v_w^2\propto
1/(r\sin\theta)^2$.  In fact, this should be a reasonably good
approximation for {\it any} radial hydromagnetic wind that has
expanded to a large distance, as the following heuristic argument
indicates.  Such winds expand more rapidly than the fast magnetosonic
velocity $c_f\equiv B/(4\pi\rho_w)^{1/2}$, where we have assumed that
thermal pressure is negligible.  At large distances, the field wraps
into a spiral with $B\simeq B_\phi$.  First, consider the flow {\it
along} streamlines.  Flux conservation gives $2\pi r\Delta r
B_\phi=$~const in radial flow.  Since $v_w$ is about constant at large
distances, it follows that $\Delta r\simeq$~const, so that
$B_\phi\propto 1/r$.  Since $\rho_w\propto 1/r^2$ in a constant
velocity, radial wind, it follows that $v_w/c_f$ is approximately
constant along a streamline at large distances.  The flow approximates
an isothermal wind, for which $(v_w/c_f)^2$ increases logarithmically
with distance from unity near the source.  This implies that at large
distances $(v_w/c_f)^2$ is approximately a constant {\it across}
streamlines as well, so that $\rho_w v_w^2 \propto B_\phi^2$. But
note that the field must become approximately force free at large
distances: balancing the tension $-B_\phi^2/(4\pi \varpi)$, where
$\varpi\equiv r\sin\theta$ is the cylindrical radius, against the
pressure gradient $-(1/8\pi)\partial B_\phi^2/ \partial\varpi$ gives
$B_\phi\propto 1/\varpi=1/r\sin\theta$.  We conclude that $\rho_w
v_w^2\propto B_\phi^2\propto 1/(r\sin\theta)^2$ is a general
characteristic of radial hydromagnetic winds.

	  This argument can be made more precise using a result due to
Ostriker (1997).  Suppose the wind arises from a Keplerian disk,
where the wind density varies with initial radius $\varpi_0$ as
$\rho_0\propto \varpi_0^{-q}$ and the Alfv\'en velocity varies with
orbital velocity at the disk.  The value $q=3/2$ corresponds to the
solution of Blandford \& Payne (1982), whereas values $0.5<q<1$ were
considered by Ostriker (1997).  Assume that each streamline expands to
$\varpi\gg\varpi_0$, so that the wind is significantly
super-Alfv\'enic.  The conservation of specific energy, angular
frequency, and mass flux along streamlines, along with the
``isorotation'' relation between $\mathbf{B}$ and $\mathbf{v}$, give
$\rho_w v_w^2\propto C(r)\varpi_0^{(1-q)/2}\varpi^{-2}$, where $C(r)$
is a slowly varying function of $r$ (Ostriker 1997).  We see that the
heuristic argument above is valid provided that the disk radius
$\varpi_0$, which varies from $\Rin$ to $\Rout$, say, has a smaller
range of variation than $\varpi$, which varies from an innermost
radius $r\theta_{\rm core}$ to $r$.  More precisely, if there is a
power law relation between $\varpi_0$ and $\sin\theta$, then
$P(\mu)\propto r^2\rho_wv_w^2$ gives
\begin{equation}\label{eq:forceindex}
P(\mu)\propto(\sin\theta)^{-2(1+\epsilon)}; ~~~
    \epsilon \simeq \left(\frac{q-1}{4}\right)
    \frac{\ln(\Rin/\Rout)}{\ln(\theta_{\rm core})},
\end{equation}
so we recover our earlier result if $\theta_{\rm core}\ll \Rin/\Rout$,
which allows the flow to be quasi-radial.  For an X-wind, $\Rin=\Rout$
and the heuristic result should be quite accurate.  Conditions at the
axis set $\theta_{\rm core}$. Although the inner boundary is generally
not considered in disk wind models, the Shu et al. (1995) theory
posits a core of open field lines from the pole of the accreting star.
The balance of magnetic pressure and tension in the fiducial X-wind
model gives $\varpi_{\rm core} = 2.5(1 + 0.18 \log_{10} r_{\pc})
~\AU$, or $\theta_{\rm core} = 1.2\times 10^{-5} r_{\pc}^{-1}
(1+0.18\log_{10} r_{\pc})$, at a distance of $r_{\pc}$ parsecs.

      We have evaluated the accuracy of equation
(\ref{eq:forceindex}) by calculating
the wind force distribution analytically using the method
outlined by Shu et al. (1995), as Ostriker (1997) suggested. We
present this calculation in Matzner (1999),
where we find that equation (\ref{eq:forceindex}) is a good
approximation at all angles more than fifteen degrees from the
equator, and best matches the actual solution between one and ten
degrees from the outflow axis. 
The fiducial X-wind model (Shu et al. 1995) has 
$P(\mu) \sin^2\theta$
greater by $40\%$ toward the equator, for reasons not considered in
equation (\ref{eq:forceindex}); this corresponds to $\epsilon \simeq
-1/50$ within $10^\circ$ of the axis at a distance of $0.1~\pc$. 

	Although the approximation
$P(\mu)\propto(\sin\theta)^{-2(1+\epsilon)}$ is valid for ideal
axisymmetric winds, we expect the force distribution to flatten within
some angle $\theta_0>\theta_{\rm core}$ in reality. This flattening
could be produced by any of the mechanisms thought to create
Herbig-Haro objects, e.g., jet precession, internal shocks from a
fluctuating wind velocity, or the magnetic kink instability.  Assuming
that $\epsilon$ is negligible and that $\theta_0\ll 1$, we can
therefore approximate the force distribution of a magnetized
protostellar wind as
\begin{equation}\label{eq:ptilde}
P(\mu)
\simeq \frac{1}{\ln(2/\theta_0)
\left(1+\theta_0^2-\mu^2\right)},
\end{equation}
where the prefactor assures $\int_0^1 P(\mu)d\mu=1$.
Outflows may require a larger
value of $\theta_0$ than appropriate for winds themselves, if mixing
between sectors (neglected here) dilutes the momentum on
axis; the current theory will still apply, with this larger
$\theta_0$.
However, so long as $\theta_0\ll 1$, the wind force is 
tightly concentrated along the axis: The formation of jets
is an inevitable consequence of a hydromagnetic wind.

\section{Shells driven by protostellar winds} \label{S:shells} 

	We shall now explore the structure and motion of a shell of
ambient material struck by a wind with the force distribution given by
equation (\ref{eq:ptilde}). Following Shu et al. (1991), Masson \&
Chernin (1992), and Li \& Shu (1996), we idealize the swept-up shell
as thin and momentum-conserving. This is justified because both shocks
are radiative for protostellar wind velocities (Koo \& McKee 1992a;
however, see \S \ref{S:conclusions} for a consideration of magnetic
pressure).  We will also adopt the assumption that the flow is
entirely radial, so that mass and momentum are conserved in each
angular sector and there is no relative motion of the shocked fluids.

A shell is driven by a ``heavy'' wind if less ambient material than
wind material has been swept up; such shells travel at nearly the wind
velocity, and the crossing time of the wind is therefore comparable to
the outflow age. Alternatively, a shell driven by a ``light'' wind is
one that has swept up more ambient gas, and has decelerated
significantly. In the limit of a very light wind, both the wind's mass
and its flight time can be neglected; this limit is approached rapidly
once a comparable mass has been swept up (Koo \& McKee 1992b).
Molecular outflows expand five to twenty times more slowly than their
driving winds, and are comparably more massive. We may therefore
neglect the wind's mass and flight time, and integrate the equation of
momentum conservation in each direction $\mu$,
$d\pw/d\Omega=v_s\partial M_a(R_s,\mu)/\partial \Omega$, where $M_a$
is the ambient mass inside the shell radius $R_s$.  We assume that the
ambient gas has a density $\rho_a=\rhoo r^{-\krho}Q(\mu)$, where
$\rhoo$ is a constant and the angular factor $Q(\mu)$ is normalized so
that $\int_0^1Q(\mu)d\mu=1$.  We find that the shell radius is
\begin{equation}\label{eq:shellKrho}
R_s^{4-\krho} = \frac{(4-\krho)(3-\krho)P(\mu)}
      {4\pi\rhoo Q(\mu)} \int_0^t \pw(t') dt'.
\end{equation}
If the wind momentum is a power law in time, $\pw\propto t^\etain$, 
then the shell velocity is given by $v(\mu) = \eta R_s/t$, where 
\begin{equation}\label{eq:eta} 
\eta\equiv \partial\ln R_s(\mu,t)/\partial\ln t =
(\etain+1)/(4-\krho).
\end{equation}
Equation (\ref{eq:shellKrho}) shows that the shell expands
self-similarly, regardless of the wind history: its radial and
velocity structures are fixed, while its scale expands as $\int\pw dt$
increases. Self-similarity is expected, because we have chosen a
scale-free medium; other radial scales, such as the scale of the
light-heavy wind transition, the wind collimation scale, and the cooling
length, are all small compared to a typical outflow. 

Shu et al. (1991), Masson \& Chernin (1992) and Li \& Shu (1996) have
all considered a steady wind ($\etain=1$) and an ambient distribution
appropriate for pre-stellar cores at the point of collapse: $\krho=2$,
and $Q(\mu)$ larger toward the equator, because of magnetic or
rotational flattening. However, this is appropriate only
in the region whose gravity is dominated by the pre-stellar core, or
where it prescribes a density higher than the ambient density: $r
\lesssim 0.07 (\sigma_{\rm th}/0.2~{\rm km~s^{-1}}) (\nH/10^4~\cm^{-3})^{-1/2}
~\pc$, using the theory of Shu (1977).
\footnote{ Note that if the wind, star and core masses are each within
about a factor of ten of the last, then the outflow must have emerged
from its core if the wind is to be light near its axis. This follows
from the overwhelming factor ($\sim 10^4$) by which the axial force is
enhanced; it is only exacerbated by any flattening of the core. The core mass
may continue to affect the low-velocity, equatorial flow, however.}

Outside this radius, anisotropies in the ambient gas are unlikely to
correlate with the outflow direction, so we may assume $Q(\mu)\simeq
1$.  We also then expect $\krho \simeq 0$, $1$ or $\gtrsim 2$, if the
lobe in question is smaller than, comparable to, or emerging from its
parent star-forming ``clump''. Assuming the ambient medium is
isotropic and the outflow has not escaped its clump, and that
$\epsilon\ll 1$ so that its effect can be neglected in the outflow
shape, our model reduces to:
\begin{equation}\label{eq:modelshape}
\frac{R_s}{R_{\rm head}} =
\left[1 + (1-\mu^2)\theta_0^{-2}\right]^{-1/(4-\krho)}, 
\end{equation} 
where the radius of the lobe head, $R_{\rm head}$, expands according
to equations (\ref{eq:ptilde}) and (\ref{eq:shellKrho}) with $\mu=1$.

\section{Comparison with observations}\label{S:obs}

	The shell described above is a Hubble flow in the sense that
${{\bf v}_s}(\mu,t) = \eta {{\bf R}_s}(\mu,t)/t$. Since the relative
line-of-sight velocity is related to the line-of-sight distance by
$v_{\rm obs} = \eta z_{\rm los}/t$, the position-velocity (PV) diagram
along the extent of an optically thin outflow is the same as its image
to an observer situated in the plane of the sky along the short axis
of the outflow. Whereas inclination causes a foreshortening of the
outflow in a sky map, it causes the PV diagram to rotate.  An
elongated outflow ($\theta_0\ll 1$) will display a nearly linear
PV diagram from the greatest to the least values of velocity and
position.  The agreement of self-similar outflows with the observed
Hubble law was pointed out by Shu et al (1991) for the particular case
$\krho=2$ and $\etain=1$; here we see that it is a quite general
property.

\begin{figure*}
\centerline{\epsfig{figure=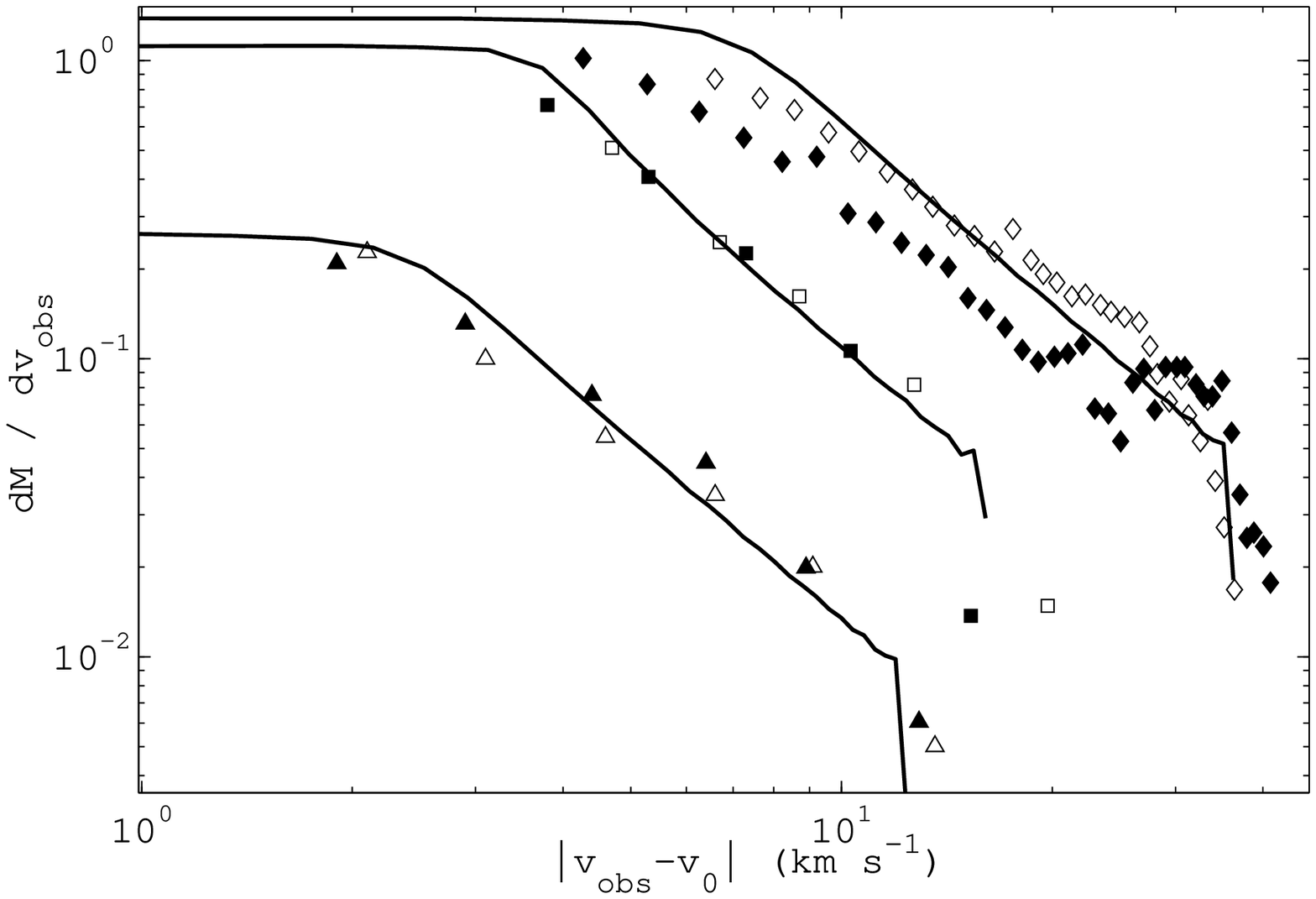,
width=4in, height = 2.2in, angle=0}}
\caption{Mass-velocity
 relations for L1551 ({\em triangles}, \cite{MS88}), NGC2071 ({\em
 squares}: \cite{MSH89}), and NGC2264G ({\em diamonds}: \cite{LF96}),
 fit with an outflow model (eq. [\ref{eq:shellKrho}]: $\krho =1$,
 $\theta_0=0.01$, $\epsilon=0$, {\em solid lines}). The inclination
 has been chosen to make the spectral break near the lowest observed
 velocities: $50^\circ$ for L1551, $40^\circ$ for NGC2071 and
 NGC2264G. {\em Filled symbols}: blue lobes; {\em open symbols}: red
 lobes. Velocities are relative to $v_0$, a center velocity, chosen to
 maximize the symmetry between each pair of lobes.  Curves have been
 offset vertically for clarity. We have taken $\epsilon=0$ so that the
 slope $\Gamma$ is $-2$ in the model; but see \S
 \ref{S:conclusions}. \label{fig:fig1}}
\end{figure*}

	 The mass-velocity relationship, $dM/dv_{\rm obs}$ is a
projection of the PV diagram onto the velocity axis.  Typical outflows
show $dM(v_{\rm obs})/dv_{\rm obs}\propto v_{\rm obs}^\Gamma$, with
$\Gamma\simeq -1.8$.  In an elongated outflow of inclination $i$,
$v_{\rm obs} \simeq v \cos i$, where $v = |{\mathbf v}|$, because all
but the lowest velocities are achieved at small angles from the
outflow axis. Therefore, $\Gamma = d\ln(dM/dv)/d\ln(v)$ except at low
velocities. Because $dM/dv= (\partial M/\partial\mu)/(\partial
v/\partial\mu) \propto v^{3-\krho}/(\mu
v^{(5+\epsilon-\krho)/(1+\epsilon)})$ 
from equations (\ref{eq:ptilde})
and (\ref{eq:shellKrho}) with $Q(\mu)=1$, we find
\begin{equation}\label{eq:Gamma}
\Gamma=  -2 + \epsilon \frac{4-\krho}{1+\epsilon}, 
\end{equation}
for $v_{\rm obs}\gtrsim 2v_{\rm min}$ where $v_{\rm min}\simeq
\theta_0^{2/(4-\krho)}v_{\rm head}$ is the minimum space
velocity 
(see eq. \ref{eq:modelshape}). 
Essentially all of the (non-equatorial) flow shows a value
of $\Gamma$ very close to $-2$ (if $\epsilon \ll 1$), in excellent
agreement with a typical observed value of $-1.8$.  The slope is
shallower for $v_{\rm obs}\lesssim v_{\rm min}$: because $dM/dv_{\rm
obs}$ is a symmetric function of $v_{\rm obs}$, $\Gamma=0$ when
$v_{\rm obs}=0$. The velocity at which $dM/dv\propto v^{-2}$ fails is
greater for more inclined outflows, or for larger $\theta_0$; 
this could in principle constrain the inclination.  The ability of our
model to fit observational mass-velocity curves is demonstrated in
Figure \ref{fig:fig1} for L1551, NGC2071, and NGC2264G.
The qualitative agreement of X-winds in isothermal toroids with
observed mass-velocity curves has previously been shown by Li \& Shu
(1996).  Again, we see that this is a general feature of
momentum--conserving shells driven by hydromagnetic winds, and is not
linked to particular models for the source of the wind nor the ambient
medium, so long as this medium has a power law density distribution.

	The broadening angle $\theta_0$ can be constrained using the
range of velocities for which $\Gamma\simeq -2$: this law holds for at
least a factor of 4 in the outflows shown in Figure 1. This must be
less than about $v_{\rm head}/(2v_{\rm min}) =
\theta_0^{-2/(4-\krho)}$, so $\theta_0 \lesssim 10^{-1.8(1-\krho/4)}$.
However, the velocity factor is also reduced by inclination at a given
$\theta_0$; in Figure 1 we show that $\theta_0 = 10^{-2}$ allows an
inclination of $\sim 45^\circ$ to fit the data;
$\theta_0 = 10^{-1.5}$ gives too small a velocity range unless $i=0$.

      If $p_{\rm obs,\,lobe}$ is the net momentum of an outflow lobe along
the line of sight, another constraint on $\theta_0$ comes from the
ratio, 
\begin{eqnarray}\label{eq:findTheta0}
\frac{ p_{\rm obs,\,lobe} } {(v_{\rm obs}^2 dM/dv_{\rm obs})_{\rm
head} }=
\frac{2\ln(\theta_0^{-1})}{4-\krho},
\end{eqnarray}
valid if $\epsilon, \theta_0\ll1$.  The result was obtained by
integrating the momentum along the line of sight for an uninclined
($i=0$) outflow; we did not assume that $\mu\simeq 1$, but instead
used the exact expression $v_{\rm obs} = \mu v$ appropriate for each
shell. It is interesting to note that this agrees exactly with the
expression $\ln(v_{\rm head}/v_{\rm min})$ one would estimate from
$dm/dv\propto v^{-2}$ with $\mu=1$.  Although the ratio was derived
for zero inclination, it is actually independent of $i$, because the
numerator and denominator scale together. Observations place a lower
limit on the ratio (an upper limit on $\theta_0$), since the net
momentum may be underestimated. We find that $\theta_0 < 10^{-1.5}$
for L1551 (\cite{MS88}) and $\theta_0 < 10^{-1.3}$ for NGC2071
(\cite{MSH89}), assuming $\krho=0$. Again, note that $\theta_0\simeq
10^{-2}$ is consistent with the data.

The current model reproduces the typical extents and velocities of
observed outflows.  For instance, equations (\ref{eq:ptilde}) and
(\ref{eq:shellKrho}) imply that a wind of momentum $20~\Msun~{\rm km~s^{-1}}$
with $\theta_0 = 10^{-2}$, blowing steadily into a uniform density
$10^4~\cm^{-3}$ for $10^5$ years, drives lobes whose heads decelerate
to $7.4~{\rm km~s^{-1}}$ and expand to $1.5~\pc$ (each) in this time.

\section{Conclusions}\label{S:conclusions}

Perhaps the most remarkable properties of bipolar molecular outflows,
apart from their intensity and frequency in regions of active star
formation, are their high degrees of collimation and the commonality
of the relation $dM/dv \propto v^{-1.8}$. We have shown that
hydromagnetic winds are naturally collimated so that the force
distribution, $\rho_w v_w^2\propto 1/\sin^2\theta$, approximately
leads to $dM/dv \propto v^{-2}$ in {\em any} power-law ambient medium,
provided the interaction is momentum conserving.  Our results show
that that the conclusions reached by Shu et al (1991) and (1995) and
Li \& Shu (1996) for steady X-winds in media with $1/r^2$ density
distributions are far more general, and in addition are in good
agreement with observations of protostellar outflows.

	 The fact that outflows are often observed to have $dM/dv$
slightly shallower than $v^{-2}$ indicates that the theory is only
approximate.  A number of effects we have not considered could
lead to a deviation from the -2 slope: CO self-absorption at lower
velocities, mixing of radial momentum between angles, or the
generation of lateral momentum when the wind impacts the shell (Masson
and Chernin 1993).  In the current model, $\Gamma$ can differ from
$-2$ either because of the wind force does not exactly follow $\rho_w
v_w^2\propto (\sin\theta)^{-2}$ ($\epsilon\neq 0$), or because the
ambient medium is not a single power law ($\krho$ varies). The model
also predicts a flattening of $dM/dv$ at low velocities, which could
raise the estimate of $\Gamma$.

Let us consider the possibility that $\epsilon$ is to blame for
$\Gamma \neq -2$. From equation (\ref{eq:Gamma}), $\Gamma\simeq -1.8$
requires $\epsilon^{-1} \simeq 5(4-\krho)$: the wind is slightly more
concentrated toward the axis.  Equation (\ref{eq:forceindex}) implies
that the number of decades of disk radius required to give this value
of $\epsilon$ is approximately $\log_{10}(1/\theta_{\rm
core})/[5(q-1)(1-\krho/4)]$.  The Blandford \& Payne (1982) model has
$q=3/2$; it would therefore require about 2-4 decades of disk radius
to give $\Gamma=-1.8$ for $\theta_{\rm core}\simeq 10^{-5}$ and
$2\geq\krho\geq 0$.  A disk with $q<1$ (initial wind density
increasing outward, e.g., Ostriker's [1997] models), has its wind
force weighted toward the equator relative to $\rho_w v_w^2\propto
(\sin\theta)^{-2}$, and produces outflows steeper than $dM/dv\propto
v^{-2}$.  X-wind models share this behavior, as they predict
$\epsilon\simeq -1/50$.

Our model assumes that the shocked wind and ambient gas form a thin
shell. Although this is appropriate for unmagnetized gas
(\cite{KM92a}), the fact that winds are collimated magnetically raises
the possibility that outflows might become inflated with a cocoon of
magnetically-supported shocked wind, before or after the wind
shuts off. This depends on the wind's terminal Alfv\'en Mach number
(inversely proportional to Poynting flux), and also on whether the
field remains ordered or becomes tangled. Because the Poynting flux
decreases as the wind collimates ($\theta_{\rm core}$ decreases),
and also as the kink instability develops (Choudhuri \& K\"onigl
1986), it is reasonable to ignore the magnetic pressure of the
shocked wind.

\acknowledgements

E. Ostriker informs us of similar, unpublished results she obtained
independently. We are grateful to F. Shu for thoughtful suggestions,
and to R. Plambeck and M. Hogerheijde for sharing and discussing their
data. CDM appreciates comments from J. Monnier.  The research
of both CDM and CFM is supported in part by the National Science
Foundation through NSF grant AST 95-30480, in part by a NASA grant to
the Center for Star Formation Studies, and, for CFM, in part by a
Guggenheim Fellowship.  CFM gratefully acknowledges the hospitality of
John Bahcall of the Institute for Advanced Study; his visit there was
supported in part by a grant from the Alfred P. Sloan Foundation.




\end{document}